\documentclass[journal]{IEEEtran}

\IEEEoverridecommandlockouts                             
\usepackage[pdftex]{graphicx}
\graphicspath{{figure/}}
\usepackage{amsmath}
\usepackage{amssymb}
\usepackage{url}

\usepackage{amsthm} 
\usepackage{cite}
\usepackage{color}
\usepackage{comment}
\usepackage{fancyhdr} 

\usepackage{algorithm}
\usepackage{algorithmic}
\usepackage{bm}

\usepackage[colorlinks=true, allcolors=blue]{hyperref}

\newtheorem{theorem}{\textbf{Theorem}}
\newtheorem{lemma}{\textbf{Lemma}}
\newtheorem{assumption}{\textbf{Assumption}}

\newcommand{\bx}{\bm{x}}
\newcommand{\by}{\bm{y}}
\newcommand{\bz}{\bm{z}}
\newcommand{\bv}{\bm{v}}
\newcommand{\bw}{\bm{w}}
\newcommand{\bd}{\bm{d}}
\newcommand{\bu}{\bm{u}}
\newcommand{\bl}{\bm{\lambda}}
\newcommand{\R}{\mathbb{R}}
\newcommand{\eS}{\mathbb{S}}
\newcommand{\D}{\Delta}

\newtheorem{corollary}{Corollary}
\newtheorem{proposition}{Proposition}

\allowdisplaybreaks

\title{\LARGE \bf Exponential Stability of Primal-Dual Gradient Dynamics with Non-Strong Convexity }

\author{ Xin Chen, Na Li
	\thanks{  X. Chen and N. Li are with the School of Engineering and Applied Sciences, Harvard University, USA. Email: (chen\_xin@g.harvard.edu, nali@seas.harvard.edu).}
	\thanks{ 
		The work was supported by
		NSF 1608509, NSF CAREER 1553407, AFOSR YIP, and ARPA-E through the NODES program.
	} 
}

\begin{document}
	
\maketitle

\thispagestyle{empty}

 \renewcommand{\headrulewidth}{0pt} 

	
\begin{abstract}
This paper studies the exponential stability of primal-dual gradient dynamics (PDGD) for solving convex optimization problems
where constraints are in the form of $A\bx+B\by=\bd$ and the objective is $\min_{\bx,\by}f(\bx)+g(\by)$
 with strongly convex smooth $f$ but only convex smooth $g$. 
We show that when $g$ is a quadratic function or when $g$ and matrix $B$ together satisfy an inequality condition, 
 the PDGD can achieve global exponential stability given that matrix $A$ is of full row rank. 
 These results indicate that the PDGD is locally exponentially stable  with respect to any convex smooth $g$ under a regularity condition. To prove the exponential stability,
two quadratic Lyapunov functions are designed. Lastly, numerical experiments  further   complement the theoretical analysis.
\end{abstract}


\section{Introduction}

Continuous-time primal-dual gradient dynamics (PDGD) \cite{int1} is a prominent first-order method to solve constrained convex optimization problems. Due to its simple structure and scalability, 
 PDGD  has been widely used in many fields, such as  wireless communication \cite{c1,c2}, power grid operation \cite{po1,na,xin}, distributed resource allocation \cite{ne1,ne2}, and imaging processing \cite{im1}. Theoretic analysis of the performance of PDGD, especially its convergence property, recently received considerable  attention.  A number of studies \cite{asy1,asy2,asy3} were devoted to establish the asymptotic stability of PDGD using local convexity-concavity of the saddle point function, while this paper focuses on a stronger stability guarantee: the global exponential stability of PDGD.

Global exponential stability is highly desired in practice. On the one hand, it is necessary to have strong stability guarantee on practical dynamic systems, especially for those in critical infrastructures like power grids and telecommunication networks. On the other hand, global exponential stability implies many useful theoretic properties. For example, using explicit Euler discretization with sufficiently small step size, a continuous-time dynamics with global exponential stability can be discretized as an iterative algorithm that achieves linear convergence rate \cite{dis1,dis2}. In addition, for a perturbed system with bounded perturbation, its solution is proved to be ultimately bounded if the corresponding nominal system is globally exponentially stable \cite{per}.

There have been a number of efforts \cite{exp1,exp2,ee,guan, ll1,ll2} in studying the exponential stability of PDGD. For equality constrained convex optimization, 
\cite{exp1,exp2} show that PDGD is globally exponentially stable when the objective is strongly convex and smooth. In \cite{ee},  the theory of integral quadratic constraints is applied to prove the global exponential stability of a proximal primal-dual flow dynamics.
Then \cite{guan} extends the global exponential stability result to the  
strongly convex optimization with  affine inequality constraints using an augmented Lagrangian function.
Besides, the local exponential stability of PDGD is established in \cite{ll1,ll2} by analyzing the spectral bounds of saddle matrices. 
As for the discrete-time counterpart, the primal-dual gradient descent algorithm solving a saddle point problem is proved to achieve linear convergence when the objective is strongly convex in the primal variables and strongly concave in the dual variables \cite{qq}. While recent work \cite{linear1,linear} shows that only one of the strong convexity (concavity) condition is necessary if  the primal-dual coupling is bilinear and the coupling matrix is of full rank.

In all the literatures above, a strongly convex and smooth objective is assumed for PDGD to achieve global exponential stability.
However, in many applications, the objective of the  optimization problem may {not} have strong convexity in all variables, such as resource allocation \cite{ne1} and optimal network flow in power systems \cite{na, xin}. Those problems can be generalized as formulation (\ref{pro})
\begin{subequations} \label{pro}
	\begin{align}
	\min_{\bm{x}\in\R^n, \by \in \R^m}&\ f(\bx)+g(\by) \label{pro-obj} \\
	s.t.\quad\ \   &A \bx +B\by = \bd
	\end{align}
\end{subequations}
where $A \in \R^{k\times n}, B\in \R^{k\times m}, \bd\in \R^k$,  $f$ is strongly convex, but $g$ is only convex (e.g. affine function or missing).

Then a natural question to ask is whether PDGD can maintain global exponential stability when the objective is non-strongly convex; if not,  what other conditions it requires to ensure at least the exponential convergence of some critical variables. 



\textbf{Contribution.} This paper 
establishes the conditions under which PDGD can still achieve global exponential stability even though the objective is non-strongly convex. 
 Specifically, we employ PDGD to solve a class of convex optimization problems in the form of (\ref{pro}), where $f$ is strongly convex smooth but $g$ is only convex smooth. Given that matrix $A$ is of full row rank, it is proved that  
 PDGD is  
globally exponentially stable when  $g$ is a quadratic function or satisfies an inequality condition together with matrix $B$.
 We further show that PDGD is 
locally exponentially stable for any convex $g$ under a  regularity condition. Two quadratic Lyapunov functions with non-zero off-diagonal terms are designed to prove these results. Lastly, we provide numerical studies to complement the analysis.

The remainder of this paper is organized as follows: Section \ref{pro_sta} introduces the detailed problem and some preliminaries. In Section  \ref{problem}, we present the main results of exponential stability, and Section \ref{proof} provides the proof sketch for these results. Numerical demonstration is carried out in Section \ref{numerical} and conclusions are drawn in Section \ref{conclusion}.

\textbf{Notations.} Throughout this paper, we 
use capital letters to denote matrices, lower case letters to denote scalars, and bold lower case letters to denote column vectors, respectively. Let $\langle \cdot , \cdot\rangle$ represent Euclidean inner product, and let $||\cdot||$ denote Euclidean norm for vectors and spectrum norm for matrices. $\R^n$ represents $n$-dimension real number space, and $\eS^{m\times m}$ denotes $m\times m$ symmetric matrix space. For two symmetric matrices $P_1$ and $P_2$, notation $P_1\succeq P_2$ means $P_1-P_2$ is positive semi-definite.

\section{Problem Statement and Preliminaries} \label{pro_sta}

In this section, we apply PDGD to solve a convex optimization problem (\ref{pro}), and present the necessary conditions for the discrete-time counterpart of PDGD to achieve linear convergence.

\subsection{Problem Statement}
This paper considers solving a class of convex optimization problems in the form of (\ref{pro}).

Define the Lagrangian function of problem (\ref{pro}) as 
\begin{align} \label{lag}
L(\bx,\by,\bl) = f(\bx)+g(\by) + \bl^\top \left( A \bx +B\by - \bd \right)
\end{align}
with dual variables $\bl\in \R^k$. Then PDGD (\ref{pdg}) is used to find the saddle points of the
Lagrangian function $L(\bx,\by,\bl)$ 
\begin{subequations} \label{pdg} 
	\begin{align}
	\dot{\bx} & = - \eta_x\cdot \nabla_{\bx} L(\bx,\by,\bl)= - \eta_x\cdot\left(\nabla f(\bx)+ A^\top \bl\right) \\
	\dot{\by}  &= - \eta_y\cdot \nabla_{\by} L(\bx,\by,\bl) = -\eta_y\cdot\left( \nabla g(\by)+B^\top\bl \right)\\
	\dot{\bl} & =  \ \ \eta_\lambda\cdot \nabla_{\bl} L(\bx,\by,\bl) = \ \ \eta_\lambda\cdot\left( A\bx+B\by-\bd\right)
	\end{align}
\end{subequations}
where $\eta_x,\eta_y,\eta_\lambda>0$ are the corresponding time constants.

We consider the case when \emph{$f(\bx)$ is strongly convex while $g(\by)$ is not} and restate this condition as assumption \ref{fa}.
\begin{assumption}\label{fa}
	Function $f$ is twice differentiable, $\mu$-strongly convex and $\ell$-smooth ($0<\mu\leq \ell$), i.e., for all $\bx_1, \bx_2\in \R^n$, 
	\begin{align}
	\begin{split}
	\mu ||\bx_1-\bx_2||^2 \leq \langle  \nabla f(\bx_1)-\nabla f(\bx_2)&,\bx_1-\bx_2\rangle \\
	&\leq \ell ||\bx_1-\bx_2||^2
	\end{split}
	\end{align}
	And function $g$ is twice differentiable, convex and $\rho$-smooth ($\rho\geq0$), i.e., for all $\by_1, \by_2\in \R^m$, 
	\begin{align}
	\begin{split}
	0 \leq \langle  \nabla g(\by_1)-\nabla g(\by_2)&,\by_1-\by_2\rangle \leq \rho ||\by_1-\by_2||^2
	\end{split}
	\end{align}
\end{assumption}

We further make the following two assumptions.
\begin{assumption} \label{finite}
	Problem (\ref{pro}) has a finite optimum.
\end{assumption}
\begin{assumption}\label{aa}
	Matrix $A$ is of full row rank and $$\kappa_1 I \preceq AA^\top \preceq \kappa_2 I$$ for some $0<\kappa_1\leq \kappa_2 $.
\end{assumption}
Note that assumption \ref{aa} is crucial for PDGD (\ref{pdg}) to achieve global exponential stability \cite{linear}, since matrix $A$ is the key connection between $\bx$, $\bl$ and $B\by$. 
By checking the KKT conditions of problem (\ref{pro}), we have the following proposition. 
\begin{proposition} 
	Under assumption \ref{fa} and \ref{finite}, any equilibrium point $(\bx^*,\by^*,\bl^*)$ of the primal-dual gradient dynamics (\ref{pdg}) is  an optimal solution of problem (\ref{pro}).
\end{proposition}
 
 \subsection{Necessary Conditions}
  
Suppose that $\bx$ is the critical decision variables that we focus on, and we aim to find out the conditions ensuring the exponential convergence of $\bx$.  
To develop some intuitions for this problem, we consider the discrete-time counterpart (\ref{pdg-d}) of PDGD (\ref{pdg}) as follows
\begin{subequations} \label{pdg-d}
	\begin{align}
	\bx_{i+1} & = \bx_i - \nu_x\cdot(\nabla f(\bx_i)+A^\top \bl_i) \label{dk-x}\\
	\by_{i+1} & = \by_i - \nu_y\cdot(\nabla g(\by_i)+B^\top \bl_i)\label{dk-y} \\
	\bl_{i+1}& = \bl_i +\nu_\lambda\cdot (A\bx_i+B\by_i-\bd) \label{dk-l}
	\end{align}
\end{subequations}
where $i$ denotes the iteration number, and $\nu_x,\nu_y,\nu_\lambda>0$ are the corresponding step sizes. 

Then the following proposition shows the convergence synchronicity of different variables and implies the necessary conditions for achieving linear convergence rate of $\bx$. 
\begin{proposition}\label{syn} 
	For the primal-dual gradient algorithm (\ref{pdg-d}), suppose that $\{\bx_i\}$ achieves linear convergence rate, in the sense that, there exists $c_x\geq 0$ and $\vartheta \in (0,1)$ such that 
	\begin{align}
	||\bx_i -\bx^*||\leq c_x \cdot \vartheta^i
	\end{align}
	Then $\{\by_i\}$ and $\{\bl_i \}$ also achieve linear convergence rate, in the sense that there exist constants $ c_\lambda,c_y,c_g\geq 0$ such that 
	\begin{subequations}
			\begin{align}
			||\bl_i -\bl^*||&\leq c_\lambda \cdot  \vartheta^i \\
		||B(\by_i -\by^*)||&\leq c_y \cdot \vartheta^i  \\
			||B\left(   \nabla g(\by_i) -\nabla g(\by^*)\right)||&\leq c_g \cdot \vartheta^i  
		\end{align}
	\end{subequations}
\end{proposition}
The proof of proposition \ref{syn} is provided in Appendix \ref{pf_syn}.
Proposition \ref{syn} indicates that the convergence of different variables is not separate but exhibits synchronicity, hence it is suggested to analyze the 
the exponential stability of PDGD (\ref{pdg}) in all variables simultaneously.

\section{Main Results } \label{problem}
In this section, we consider the quadratic case and the general case of $g(\by)$, and
present the global (local) exponential stability results of PDGD (\ref{pdg}). 

For  explicit expression, we stack $\bx$, $\by$, $\bl$ into  vector $\bz := \left[ \bx^\top, \by^\top, \bl^\top\right]^\top$ and define $\bz^* := \left[ \bx^{*\top}, \by^{*\top}, \bl^{*\top}\right]^\top$ as one of the equilibrium points of PGDG (\ref{pdg}).

\subsection{Quadratic Case of $g(y)$}
Consider the quadratic case when $g(\by)$ is a quadratic function given by
\begin{align}\label{quag}
g(\by)=\frac{1}{2}\by^\top G\by+\bm{g}^\top \by +g_0
\end{align} 
 with $g_0\in \R$, $\bm{g}\in \R^m$, $G\in \eS^{m\times m}$.  By assumption \ref{fa}, we have $0\preceq G\preceq \rho I$.
 
 For this case, the equilibrium point set $\Psi$ of PDGD (\ref{pdg}) is specified by proposition \ref{eps-q}, and its global exponential stability is stated as theorem \ref{QUA}.
 
\begin{proposition} \label{eps-q} Under assumption \ref{fa}, \ref{finite} and \ref{aa}, when $g(\by)$ is a quadratic function in the form of (\ref{quag}), the  equilibrium point set $\Psi$ of the primal-dual gradient dynamics (\ref{pdg}) is given by
	\begin{align} \label{eqs-qua}
	\Psi := \left.\{ \hat{\bz} | \ \hat{\bx} = \bx^*, \hat{\bl} = \bl^*, B\hat{\by}=B\by^*, G\hat{\by}=G\by^*  \right.\}
	\end{align}
\end{proposition}
See Appendix \ref{pf-epsq} for the proof of proposition \ref{eps-q}.

Proposition \ref{eps-q} implies that the components $\bx^*$ and $\bl^*$ of the equilibrium points are unique, while $\by^*$ is  non-unique and
$\hat{\by}-\by^*\in ker(B)\cap ker(G)$ for any $\hat{\bz}\in \Psi$.

\begin{theorem} \label{QUA}
	Under assumption \ref{fa}, \ref{finite} and \ref{aa}, when $g(\by)$ is a quadratic function in the form of (\ref{quag}), 
	the prime-dual gradient dynamics (\ref{pdg})   is globally exponentially stable in the sense that, there exist constants $a_x,a_\lambda,a_{y_B}, a_{y_G} \geq 0$ and $\tau >0$ such that 
	\begin{subequations} \label{ccon2}
		\begin{align}
		||\bx(t)-\bx^*|| &\leq a_x \cdot e^{-\tau t}
		\\
				||\bl(t)-\bl^*|| &\leq a_\lambda \cdot e^{-\tau t}\\
					||B(\by(t)-\by^*)|| &\leq a_{y_B} \cdot e^{-\tau t}  \label{by}\\
		||G(\by(t)-\by^*)|| &\leq a_{y_G} \cdot e^{-\tau t} \label{gy}
		\end{align}
	\end{subequations} 	
\end{theorem}

The proof of theorem \ref{QUA} is provided in Section \ref{prof-qua}. Equation (\ref{ccon2}) indicates that the distance between the solution $\bz(t)$ and $\Psi$ converges to zero exponentially.


\emph{Remark 1:} In theorem \ref{QUA}, we only take $g(\by)$ as  quadratic function, 
while $f(\bx)$ can be any strongly convex function. In this case, PDGD (\ref{pdg}) is not necessarily a linear time-invariant system, but always 
 preserves global exponential stability.

\subsection{General Case of $g(y)$}

Now  consider a general convex function $g(\by)$. Since it is not easy to analyze the global exponential stability of PDGD (\ref{pdg}) with an unclear equilibrium point set of $\by^*$, we supplement assumption \ref{ge-psd} to make optimal $\by^*$ to be unique.
{\color{black}  An intuition  for assumption \ref{ge-psd} is that if  $B$  is the all-zero matrix, assumption \ref{ge-psd} reduces to the condition that $g(\by)$ is strongly convex. Actually, assumption \ref{ge-psd} ``mimics" a strong convexity condition in $\by$ and use matrix $B$ to make up for the strong convexity deficit of $g(\by)$.
\begin{assumption} \label{ge-psd}
	For any $\by_1,\by_2\in \R^m$, there exists constant $\gamma>0$ such that 
	\begin{align} \label{mo}
	\begin{split}
	&(\by_1-\by_2)^\top B^\top B (\by_1-\by_2)
	\\&\quad + \langle  \nabla g(\by_1)-\nabla g(\by_2),\by_1-\by_2\rangle   \geq \gamma \cdot ||\by_1-\by_2||^2
	\end{split}
	\end{align}
\end{assumption}
\begin{proposition} \label{eqs-ge}
	Under assumption  \ref{fa}, \ref{finite}, \ref{aa}, \ref{ge-psd}, the prime-dual gradient dynamics (\ref{pdg}) has a unique equilibrium point $\bz^*$.
\end{proposition}
See Appendix \ref{pf-epsq} for the proof of proposition \ref{eqs-ge}. Noted that assumption \ref{ge-psd} is a sufficient condition for the uniqueness of $\bz^*$ but not necessary. For example, when $B$ is the all-zero matrix and $g(\by): = \sum_{i=1}^{m} y_i^4$, assumption \ref{ge-psd} does not hold, while $\bz^*$ is unique with $\by^* = \bm{0}$.

Accordingly, the global exponential stability of PDGD (\ref{pdg}) is established as the following theorem.

\begin{theorem} \label{thm-psd}
	Under assumption \ref{fa}, \ref{finite}, \ref{aa} and \ref{ge-psd}, the prime-dual gradient dynamics (\ref{pdg}) is globally exponentially stable, in the sense that, there exist constants $c_z\geq 0$ and $\tau >0$ such that
		\begin{align} \label{exp-1}
		||\bz(t)-\bz^*|| &\leq c_z \cdot e^{-\tau t}
		\end{align}
\end{theorem}
The proof of theorem \ref{thm-psd} is provided in Section \ref{prof-psd}.

One special case of assumption \ref{ge-psd} is that $B^\top B\succeq \gamma I$ for some $\gamma>0$, i.e., $B^\top B $ is positive definite. This  is equivalent to the condition that  matrix $B$ is of full column rank. Thus we have the following corollary.
\begin{corollary} \label{cor-1}
	Under assumption \ref{fa}, \ref{finite} and \ref{aa},  \textit{if matrix $B$ is of full column rank}, the prime-dual gradient dynamics (\ref{pdg})  is globally exponentially stable in the sense of (\ref{exp-1}). 
\end{corollary}

\subsection{Local Exponential Convergence}

For a finite-dimension nonlinear system, it is well-known that if the linearized system  based on an equilibrium point is exponentially stable, then the original system is locally exponentially stable around this equilibrium point. 
Moreover, under assumption \ref{fa} and \ref{finite}, it proves that PDGD (\ref{pdg}) globally asymptotically converges to one of the equilibrium points in $\Psi$ \cite{na,asy1}.
Inspired by those facts and the quadratic case, we claim the local exponential convergence of PDGD (\ref{pdg}) with the following theorem.  
\begin{theorem}\label{local}
		Under assumption \ref{fa}, \ref{finite} and  \ref{aa}, suppose that the trajectory $\bz(t)$ following
		the prime-dual gradient dynamics (\ref{pdg}) globally asymptotically converges to the equilibrium point $\bz^*$, if we have 
		\begin{align} \label{loo}
		     B^\top B + \nabla^2g(\by^*) \succ 0
		\end{align}
	    then there exist a time $t_\delta\geq0$
		  and constants $\tau >0$ such that for any time $ t\geq t_\delta$,
		  \begin{align}
		     ||\bz(t)-\bz^*|| &\leq ||\bz(t_\delta)-\bz^*|| \cdot e^{-\tau (t-t_\delta)}
		  \end{align} 
\end{theorem}

See Appendix \ref{pf-local} for the proof.

\section{Exponential Stability Analysis} \label{proof}

In this section, we 
present the proofs for theorem \ref{QUA} and \ref{thm-psd}. 
To begin with, we introduce the following lemma, whose proof can be found in \cite[Appendix D]{guan}.
\begin{lemma} \label{F-G}
	Under assumption \ref{fa}, for any $\bx\in \R^n$, there exists a symmetric matrix $F(\bx)\in \eS^{n\times n}$ that depends on $\bx$ and satisfies $\mu I \preceq F(\bx)\preceq \ell I$, such that $$\nabla f(\bx) -\nabla f(\bx^*) = F(\bx)(\bx-\bx^*)$$
	For any $\by\in \R^m$, there exists a symmetric matrix $G(\by)\in \eS^{m\times m}$ that depends on $\by$ and satisfies ${0} \preceq G(\by)\preceq \rho I$, such that $$\nabla g(\by) -\nabla g(\by^*) = G(\by)(\by-\by^*)$$ 
\end{lemma}

\subsection{Proof of Theorem \ref{QUA}}   \label{prof-qua}

For matrix $G$, pick up $\{\sigma_i,\bu_i\}_{i=1,2,\cdots,m}$ as its eigen-pairs and satisfying the following three properties:
\begin{itemize}
	\item [(1)] $\{\bu_i\}_{i=1,2,\cdots,m}$ form an orthonormal basis of $\R^m$.
	\item [(2)] The first $l$ eigenvalues are positive, i.e. $\sigma_1,\sigma_2\cdots,\sigma_l>0$ and $\sigma_{l+1},\sigma_{l+2},\cdots,\sigma_m =0$. 
	\item [(3)] $ker(G)\cap ker(B) = span(\bu_{l+r+1},\bu_{l+r+2},\cdots,\bu_m)$ for a certain $r\in\{0,1,\cdots,m-l\}$, where $r =m-l $ means  $ker(G)\cap ker(B) =\{\bm{0}\}$.
\end{itemize}
It can be checked that such eigen-pairs $\{\sigma_i,\bu_i\}$ always exist.

Define matrix $U\in \R^{m\times {(l+r)}}$ as 
\begin{align*}
U:= \left[\bu_1,\bu_2,\cdots,\bu_{l+r} \right]
\end{align*}
which collects all the eigenvectors of $G$ except those in the space $ker(G)\cap ker(B)$. 
Then matrix $G$ can be rewritten as $G = U\Sigma U^\top$ with $\Sigma :=\text{diag}(\sigma_1,\sigma_2,\cdots,\sigma_{l+r}) $. Since we have $row(B) \subseteq Col(U)$ by definition, there exists  a matrix $T\in \R^{(l+r)\times k}$ such that $B^\top = U T$, i.e., $B = T^\top U^\top$.

 To prove theorem \ref{QUA}, we 
design the quadratic Lyapunov function $V_1(\bz)$ as
\begin{align} \label{V1}
V_1(\bz) = (\bz -\bz^*)^\top  P_1 (\bz-\bz^*)
\end{align}
where $P_1\in \R^{(m+n+k)\times(m+n+k)}$ is defined by
\begin{align} \label{P1}
P_1 = \begin{bmatrix}
\frac{\alpha}{\eta_x} I & 0 & \frac{1}{\eta_\lambda}A^\top \\ 0 &  \frac{\alpha}{\eta_y}UU^\top &  - \frac{\beta}{\eta_\lambda}B^\top \\\frac{1}{\eta_\lambda}A & -\frac{\beta}{\eta_\lambda}B & \frac{\alpha}{\eta_\lambda} I
\end{bmatrix}
\end{align}
Here, parameter $\alpha$ is a sufficiently large positive number, and parameter $\beta$ is a sufficiently small positive number.

\begin{lemma}  \label{Psd}
	$P_1$ is positive semidefinite.  $V_1(\bz) =0$ if and only if $\bx=\bx^* $, $\bl=\bl^*$, $B\by=B\by^*$ and $G\by=G\by^*$.
\end{lemma}
See Appendix \ref{pf-Psd} for the proof of lemma \ref{Psd}.

If we can show that the time derivative of $V_1(\bz)$ along the trajectory of PDGD (\ref{pdg}) satisfies
\begin{align} \label{dV}
\frac{d V_1(\bz)}{dt} \leq -\tau V_1(\bz)
\end{align}
for $\tau = \frac{\beta^2}{\alpha}>0$, then theorem \ref{QUA} is proved. The following part is devoted to prove the property (\ref{dV}).

With lemma \ref{F-G}, PDGD (\ref{pdg}) can be equivalently rewritten as 
\begin{align} \label{dddd}
\begin{split}
\frac{d \bz}{d t} & = 
\begin{bmatrix} 
- \eta_x \left( \nabla_{\bx} L(\bx,\by,\bl) -\nabla_{\bx} L(\bx^*,\by^*,\bl^*)  \right) \\
-\eta_y\left(\nabla_{\by} L(\bx,\by,\bl)-\nabla_{\by} L(\bx^*,\by^*,\bl^*) \right)\\
\ \   \eta_\lambda\left(\nabla_{\bl} L(\bx,\by,\bl)-\nabla_{\bl} L(\bx^*,\by^*,\bl^*) \right)
\end{bmatrix} \\
& =\underbrace{ \begin{bmatrix}
-  \eta_x  F(\bx) & 0 & - \eta_x A^\top \\ 0 & - \eta_y G(\by) & - \eta_y B^\top \\  \eta_\lambda A &\eta_\lambda B & 0
\end{bmatrix}}_{: = W(\bz)} (\bz -\bz^*) 
\end{split}
\end{align}
Here, since $g(\by)$ is a quadratic function in the form of (\ref{quag}), we have 
$G(\by) \equiv G$.

Then $\frac{d V_1(\bz)}{dt}$ can be formulated as 
\begin{align}
\begin{split}
&\frac{d V_1(\bz)}{dt}  =  \dot{\bz} ^\top  P_1 (\bz-\bz^*) +  (\bz -\bz^*)^\top  P_1 \dot{\bz}\\
 &\ \ \ \ \ = (\bz-\bz^*)^\top\left[W(\bz)^\top P_1 + P_1 W(\bz)    \right]  (\bz-\bz^*)
\end{split}
 \end{align}
Hence, it is sufficient to show  property (\ref{dV}) by the following lemma. See Appendix \ref{pf-WP} for the proof of lemma \ref{WP}.
\begin{lemma} \label{WP}
   For any $\bz\in \R^{n+m+k}$, we have 
	\begin{align} \label{WP1+}
 W(\bz)^\top  P_1+	P_1  W(\bz) \preceq -\tau P_1
	\end{align}
\end{lemma}

In this way, we prove theorem \ref{QUA} using (\ref{dV}) and lemma \ref{WP}.

\subsection{Proof of Theorem \ref{thm-psd}}\label{prof-psd}

To prove theorem \ref{thm-psd}, we design the quadratic Lyapunov function $V_2(\bz)$ as
\begin{align}\label{V2}
V_2(\bz) = (\bz -\bz^*)^\top  P_2 (\bz-\bz^*)
\end{align}
where $P_2\in \R^{(m+n+k)\times(m+n+k)}$ is defined by
\begin{align} \label{P2}
P_2 = \begin{bmatrix}
\frac{\alpha}{\eta_x} I & 0 & \frac{1}{\eta_\lambda}A^\top \\ 0 &  \frac{\alpha}{\eta_y} I &  - \frac{\beta}{\eta_\lambda}B^\top \\\frac{1}{\eta_\lambda}A & -\frac{\beta}{\eta_\lambda}B & \frac{\alpha}{\eta_\lambda} I
\end{bmatrix}
\end{align}
Here, parameter $\alpha$ is a sufficiently large positive number, and parameter $\beta$ is a sufficiently small positive number.
\begin{lemma} \label{Ppd}
	$P_2$ is positive definite. $V_2(\bz) =0$ if and only if $\bz=\bz^*$.
\end{lemma}
See Appendix \ref{pf-Ppd} for the proof of lemma \ref{Ppd}.

Similar to (\ref{dV}), if we can prove that the time derivative of $V_2(\bz)$ along the trajectory of PDGD (\ref{pdg}) satisfies
\begin{align} \label{dV2}
\frac{d V_2(\bz)}{dt} \leq -\tau V_2(\bz)
\end{align}
for $\tau = \frac{\beta^2}{\alpha}>0$, then theorem \ref{thm-psd} is proved. To show property (\ref{dV2}), 
it is sufficient to prove lemma \ref{WP2}.  See Appendix \ref{pf-WP2} for its proof.

\begin{lemma} \label{WP2}
	For any $\bz\in \R^{n+m+k}$, we have 
	\begin{align} \label{WP2+}
	W(\bz)^\top  P_2+	P_2  W (\bz) \preceq -\tau P_2
	\end{align}
\end{lemma}

By lemma \ref{WP2} and (\ref{dV2}), we prove theorem \ref{thm-psd}.

\section{Numerical Examples} \label{numerical}

\subsection{Quadratic Case of $g(y)$}
For problem (\ref{pro}), let $n=60$, $m=50$ and $k=20$. The time constant is set as $\eta =\eta_x =\eta_y =\eta_\lambda$. 
Define $f(\bx) = \frac{1}{2} \bx^\top F \bx$ where $F := 5I + F_0^\top F_0$ and $F_0$ is a $n\times n$ random matrix. Define $g(\by) =  \frac{1}{2} \by^\top G \by$ where $G := \text{diag}(0,G_0^\top G_0) $ and $G_0$ is a $r_G \times (m-1)$ random matrix with $r_G = 40$; then $G$ is a positive semi-definite matrix with the rank at most $r_G$. 
 Let $B := [\bm{0},B_0]$ and 
$B_0$ is a $k\times (m-1)$ random matrix. 
Under this setting, we have $\bm{e}_1\in ker(G)\cap ker(B) $ where $\bm{e}_1 = [1,0,\cdots,0]^\top $. $A$ and $\bm{b}$ are also random matrix and random vector respectively.

\begin{figure}[thpb]
	\centering
	\includegraphics[scale=0.201]{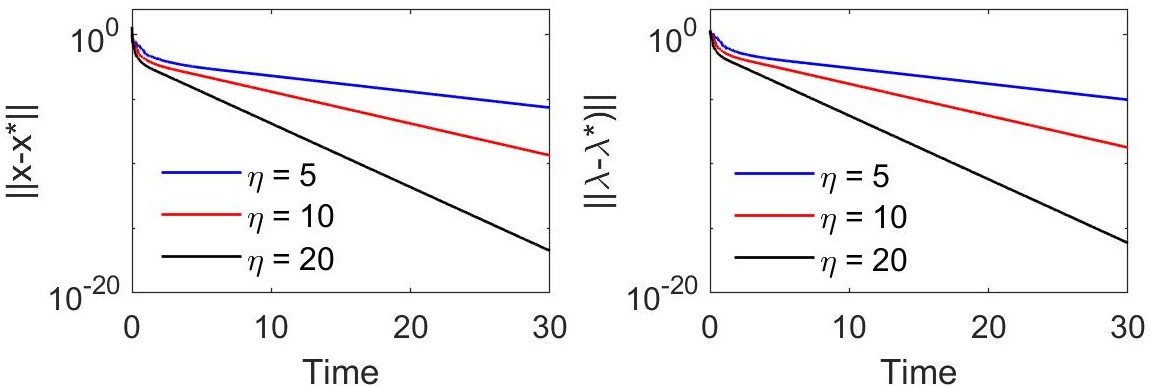}
	\includegraphics[scale=0.201]{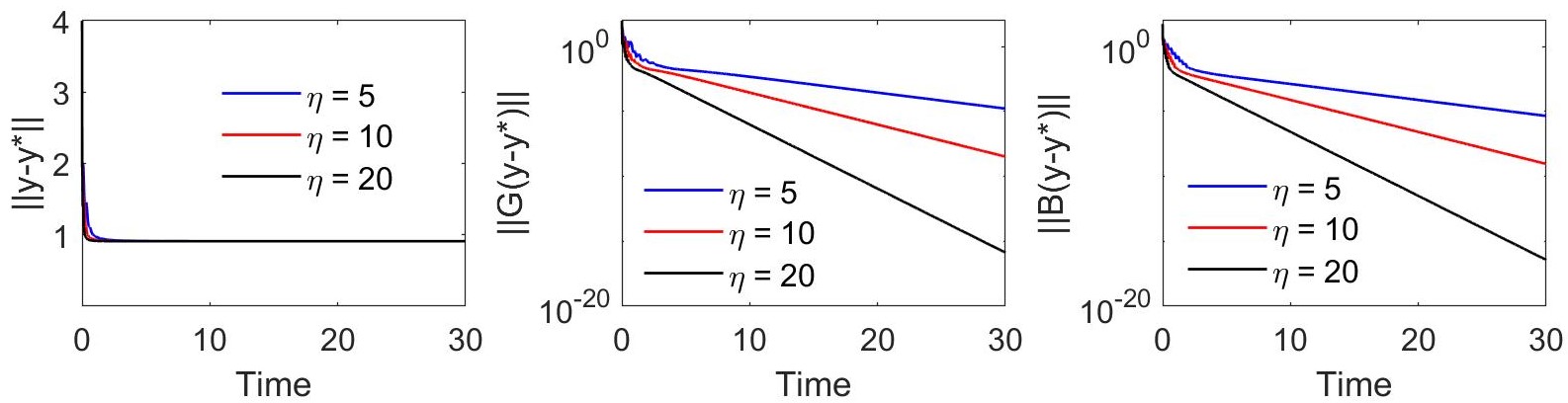}
	\caption{Convergence results of PDGD with different time constants when $g(\by)$ is a quadratic function.}
	\label{Qua_f}
\end{figure}

We pick up an arbitrary equilibrium point in $\Psi$ as $\bz^*$, then
the convergence results of PDGD are shown as figure \ref{Qua_f}. It is observed that $||\bx-\bx^*||$, $||\bl-\bl^*||$, $||B(\by-\by^*)||$ and $||G(\by-\by^*)||$   converge exponentially to zero while $||\by-\by^*||$ does not, which follows the statements in theorem \ref{QUA}.

\subsection{General Case of $g(y)$} \label{num-gen}
In this case, let $g(\by) = \sum_{i=1}^{m} y_i^4$ and $n=60$, $m=20$. $F_0, A, B, b$ are all random matrices or vector. We set $k=10$ and $k=30$ respectively, and run simulations for these two cases. The convergence results of PDGD are presented as figure \ref{Gee_f}. When $k =30$, matrix $B$ is of full column rank, thus assumption \ref{ge-psd} holds and PDGD is globally exponentially stable. When $k=10$, assumption \ref{ge-psd} is not satisfied, so PDGD exhibits asymptotic convergence at first, then (Time$>30$) exponentially converges to the equilibrium point due to the local exponential stability.

\begin{figure}[thpb]
	\centering
	\includegraphics[scale=0.21]{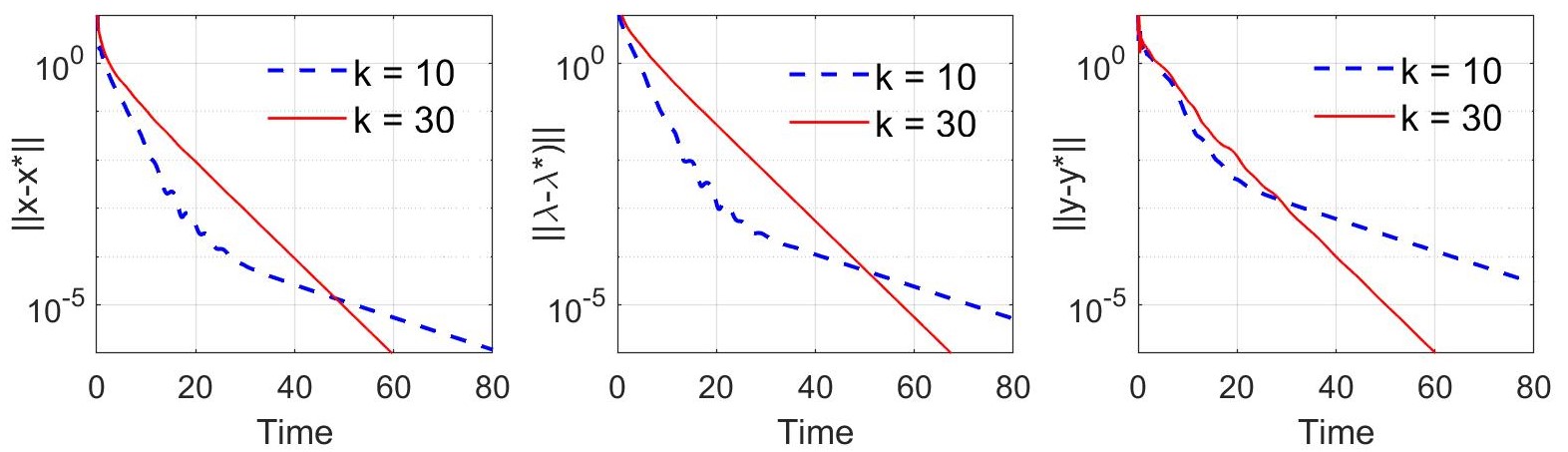}
	\caption{Convergence results of PDGD when $g(\by) = \sum_{i}^m y_i^4$.}
	\label{Gee_f}
\end{figure} 
 
\section{Conclusion}\label{conclusion}

In this paper, we prove that the primal-dual gradient dynamics (\ref{pdg}) can achieve global (local) exponential stability for convex smooth optimization problems in the form of (\ref{pro}) with non-strong convexity in their objectives. 
Our main results are summarized as theorem \ref{QUA}, theorem \ref{thm-psd} and theorem \ref{local}, which are proved with two quadratic Lyapunov functions. Numerical experiments are implemented and further validate these results.
Future work is to extend these  results to convex optimization problems with both equality and inequality constraints.

\section*{Appendix}

\subsection{Proof of Lemma \ref{WP}} \label{pf-WP}

Plugging $P_1$ (\ref{P1}) into equation (\ref{WP1+}) with $\tau =\frac{\beta^2}{\alpha}$, we obtain
\begin{align} \label{Q1z}
\begin{split}
&Q_1(\bz) := -W(\bz)^\top  P_1-	P_1  W (\bz) -\tau P_1 =\\
& \begin{bmatrix}
Q_{x}& (\beta-1)A^\top B &  \frac{\eta_x}{\eta_\lambda}FA^\top - \frac{\beta^2}{\alpha \eta_\lambda} A^\top \\
(\beta-1)B^\top A & Q_y & \frac{\beta^3}{\alpha \eta_\lambda}B^\top - \frac{\beta\eta_y}{\eta_\lambda}GB^\top\\
\frac{\eta_x}{\eta_\lambda}AF-\frac{\beta^2}{\eta_\lambda \alpha}A& \frac{\beta^3}{\alpha \eta_\lambda}B- \frac{\beta\eta_y}{\eta_\lambda}BG &Q_\lambda
\end{bmatrix} 
\end{split}
\end{align}
with 
\begin{subequations}
	\begin{align}
	&Q_{x} := 2\alpha F-2A^\top A -\frac{\beta^2}{\eta_x}I\\
	\begin{split}
	&Q_{y} := 2\alpha G + 2\beta B^\top B 
	-\frac{\beta^2}{\eta_y}UU^\top
	\end{split} \label{Qy}
 \\
	&Q_{\lambda}:=  2\frac{\eta_x}{\eta_\lambda}AA^\top -2 \frac{\beta\eta_y}{\eta_\lambda}BB^\top -\frac{\beta^2}{\eta_\lambda} I
	\end{align}	
\end{subequations}

Here, we denote $F(\bx)$ and $G(\by)$ as $F$ and $G$ respectively for  notation simplification.
In (\ref{Q1z}), we apply $B = T^\top U^\top$  (Section \ref{prof-qua}) and thus have
$$BUU^\top = T^\top U^\top UU^\top = T^\top U^\top =B$$
In (\ref{Qy}), we use
 $ GUU^\top +UU^\top G = 2{U}\Sigma {U}^\top =2G$.
 
To prove Lemma \ref{WP}, it is sufficient to show $Q_1(\bz)\succeq 0$ for any $\bz \in \R^{n+m+k}$. In the follows, we will use the Schur complement argument twice to  prove it.

Denote $\omega$ as the  largest  eigenvalue of $B^\top B$, then $0\preceq BB^\top \preceq \omega I$ and $0\preceq B^\top B \preceq \omega I$, and we have 
\begin{align} \label{Ql1}
Q_{\lambda}\succeq  \underbrace{ \left( 2\frac{\eta_x}{\eta_\lambda}\kappa_1  -2 \frac{\eta_y}{\eta_\lambda}\beta\omega  - \frac{\beta^2}{\eta_\lambda}\right)}_{:=c_1(\beta)} \cdot I 
\end{align}

Firstly, consider the  Schur complement\footnote{ For matrix $M=\begin{bmatrix}
	A& B\\ C& D
	\end{bmatrix}$, the  Schur complement of the block $D$ in $M$ is defined as $M/D: = A-BD^{-1}C$.
} of the block $Q_\lambda $ in ${Q}_1(\bz)$ (\ref{Q1z}), which is denoted as $\bar{Q}_1(\bz)$. Using (\ref{Ql1}), we have 
\begin{align} \label{hQ1}
\begin{split}
\bar{Q}_1(\bz) &\succeq \hat{Q}_1(\bz):= \begin{bmatrix}
Q_x-M_1 & M_2\\
M_2^\top  & Q_y -M_3
\end{bmatrix} 
\end{split}
\end{align}
where 
\begin{align*}
M_1 &:=  \frac{1}{c_1\eta_\lambda^2} (  \eta_xF - \frac{\beta^2}{\alpha} I)A^\top A (\eta_xF-\frac{\beta^2}{\alpha}I)
\\
M_2 &:=  (\beta-1)A^\top B -\frac{\beta}{c_1\eta_\lambda^2} (  \eta_xF - \frac{\beta^2}{\alpha} I)A^\top B(\frac{\beta^2}{\alpha}I-\eta_y G) \\
M_3 &:=  \frac{\beta^2}{c_1\eta_\lambda^2}(
\frac{\beta^2}{\alpha}I-\eta_y G
)
B^\top B(\frac{\beta^2}{\alpha}I-\eta_y G) 
\end{align*}

Since $\mu I \preceq F \preceq \ell I$ and $A^\top A \preceq \kappa_2 I$, we have
\begin{align}\label{QxM1}
\begin{split}
Q_x &-M_1  = 2\alpha F-2A^\top A-\frac{\beta^2}{\eta_x}I-M_1\\
&\succeq \underbrace{\left[ 2\alpha \mu  - 2\kappa_2   - \frac{\beta^2}{\eta_x} -
\frac{\kappa_2}{c_1\eta_\lambda^2}(\eta_x \ell + \frac{\beta^2}{\alpha})^2  \right]}_{:=c_2(\alpha,\beta)}\cdot I
\end{split}
\end{align}

Secondly, consider the Schur complement of the block $Q_x-M_1$ in $\hat{Q}_1(\bz)$ (\ref{hQ1}), which is denoted as $\tilde{Q}_1(\bz)$. Using (\ref{QxM1}), we have 
\begin{align} \label{Q1zs}
\tilde{Q}_1(\bz) \succeq {Q}_1^*(\bz):=Q_y-M_3 - \frac{1}{c_2} M_2^\top M_2
\end{align}

We claim the following lemma to show ${Q}_1^*(\bz)\succeq 0$, whose proof is provided in Appendix \ref{pf_Q_Z}.
\begin{lemma} \label{Q_Z}
	When $\alpha$ is large enough and $\beta$ is positively small enough,
	${Q}_1^*(\bz)$ in (\ref{Q1zs}) is positive semidefinite, i.e. $  {Q}_1^*(\bz)\succeq 0$, for any $\bz\in \R^{n+m+k}$.
\end{lemma}

In addition, to make the above Schur complement argument work, it requires that $c_1(\beta)>0$ and $ c_2(\alpha,\beta) >0$. This condition can be achieved if we set parameter $\alpha$ large enough and $\beta$ positively small enough. Because for $\alpha,\beta>0$,  $c_2(\alpha,\beta)$ is a scalar function that is strictly increasing in $\alpha$ and $c_2\to +\infty$ as $\alpha\to+\infty$ with fixed $\beta$;
$c_1(\beta)$ is a strictly decreasing function in $\beta$, and $c_1 \to \frac{2\eta_x}{\eta_y}\kappa_1>0$ as $\beta\to 0$. 

According to the Schur complement theorem and lemma \ref{Q_Z}, we obtain $\tilde{Q}_1(\bz)\succeq 0$, $\hat{Q}_1(\bz)\succeq 0$, $\bar{Q}_1(\bz)\succeq 0$ and eventually ${Q}_1(\bz)\succeq 0$.

\subsection{Proof of Lemma \ref{WP2}} \label{pf-WP2}
Since the proof of lemma \ref{WP2} is very similar to the proof of lemma \ref{WP},  without causing any confusion, we recycle the notations and definitions used in Appendix \ref{pf-WP}. 

 Plugging $P_2$ (\ref{P2}) into equation (\ref{WP2+}) with $\tau =\frac{\beta^2}{\alpha}$, we denote 
\begin{align}
Q_2(\bz) := -W(\bz)^\top  P_2-	P_2  W(\bz) -\tau P_2 
\end{align}
The detailed formulation of $Q_2(\bz)$ is exactly the same as $Q_1(\bz)$ (\ref{Q1z}) except that the block $Q_{y}$ in $Q_2(\bz)$ is defined as  
\begin{align}
Q_{y}^\prime := 2\alpha G+2\beta B^\top B -\frac{\beta^2}{\eta_y} I
\end{align}
Here, we use notation $Q_{y}^\prime$ to distinguish with the corresponding block $Q_{y}$ in $Q_1(\bz)$ (\ref{Q1z}).

To prove Lemma \ref{WP2}, it is sufficient to show $Q_2(\bz)\succeq 0$ for any $\bz \in \R^{n+m+k}$, and we will use the Schur complement argument twice to  prove it.

Firstly, consider the Schur complement of the block $Q_\lambda $ in ${Q}_2(\bz)$ (\ref{Q1z}), which is denoted as $\bar{Q}_2(\bz)$. Using (\ref{Ql1}), we have 
\begin{align} \label{hQ2}
\begin{split}
\bar{Q}_2(\bz) &\succeq \hat{Q}_2(\bz):= \begin{bmatrix}
Q_x-M_1 & M_2\\
M_2^\top  & Q_y^\prime -M_3
\end{bmatrix} 
\end{split}
\end{align}
Here, $Q_x$, $M_1$, $M_2$ and $M_3$ have exactly the same definitions as those in (\ref{hQ1}).

Secondly, consider the Schur complement of the block $Q_x-M_1$ in $\hat{Q}_2(\bz)$ (\ref{hQ2}), which is denoted as $\tilde{Q}_2(\bz)$. Using (\ref{QxM1}), we have 
\begin{align} \label{Q1zs2}
\tilde{Q}_2(\bz) \succeq {Q}_2^*(\bz):=Q_y^\prime-M_3 - \frac{1}{c_2} M_2^\top M_2
\end{align}

By assumption \ref{ge-psd}, we have $$2\alpha G+2\beta B^\top B \succeq 2\beta(G+B^\top B)\succeq 2\beta \gamma I$$ and thus
\begin{align}
Q_y^\prime -M_3
&\succeq \beta\cdot \underbrace{\left[  2\gamma -\frac{\beta}{\eta_y} - \frac{\beta \omega}{c_1\eta_\lambda^2} (\frac{\beta^2}{\alpha }+ \eta_y\rho )^2\right]}_{:=c_3(\alpha,\beta)} \cdot I 
\end{align}

Similar to lemma \ref{Q_Z}, when  set parameter $\alpha$ larger enough and $\beta$ positively small enough, we have ${Q}_2^*(\bz)\succeq 0$ for any $\bz\in \R^{n+m+k}$. 
The proof sketch is that for sufficiently large $\alpha$ and sufficiently small $\beta$,  $||M_2^\top M_2||$ is bounded. Since $c_2$ is dominated by $\alpha$, we can set $\alpha$ large enough such that $ \frac{1}{c_2}\leq \beta^2$ for any fixed $\beta$, then select $\beta$ positively small enough such that $c_3 -\beta||M_2^\top M_2||>0 $. 
Thus we have ${Q}_2^*(\bz)\succeq 0$.

Consequently, by the Schur complement theorem, we have $\tilde{Q}_2(\bz)\succeq 0$, $\hat{Q}_2(\bz)\succeq 0$, $\bar{Q}_2(\bz)\succeq 0$ and eventually ${Q}_2(\bz)\succeq 0$  for any $\bz \in \R^{n+m+k}$.

\subsection{Proof of Lemma \ref{Q_Z}} \label{pf_Q_Z}

As shown in equation (\ref{Q1zs}), $Q_1^*(\bz)$ is defined as 
\begin{align*}
Q_1^*(\bz):=Q_y-M_3 - \frac{1}{c_2} M_2^\top M_2
\end{align*}
Consider the three items in $Q_1^*(\bz)$ one by one as follows.

For the first term, there exists a constant $\pi>0$ such that
\begin{align*}
G + B^\top B = U\left(\Sigma + TT^\top    \right)U^\top  \succeq \pi\cdot UU^\top
\end{align*}
Because $ker(G)\cap ker(B) = span(\bu_{l+r+1},\cdots,\bu_m)$, we have $D := \Sigma + TT^\top  \succ 0$. Otherwise, if $D$ is just positive semi-definite, there exists $\bw\in \R^{l+r} \neq \bm{0}$ such that $D\bw = \bm{0}$. Then we find a vector $\bv := U\bw\neq \bm{0}$ such that 
\begin{align*}
\bv^\top (G+B^\top B) \bv = \bv^\top U D U^\top \bv = \bw^\top D \bw =0
\end{align*}
Since $G$ and $B^\top B$ are both positive semi-definite, we have 
$\bv \in ker(G)\cap ker (B)$. However, by the definition $\bv := U\bw\in span (\bu_1,\cdots,\bu_{l+r})$, which is contradictory.
Thus $D$ is positive definite and  $D  \succeq \pi I $ where $\pi$ is the smallest eigenvalue of $D$. 
As a result, we have 
\begin{align*}
\begin{split}
Q_y & =2\alpha G + 2\beta B^\top B 
-\frac{\beta^2}{\eta_y}UU^\top \succeq \beta\underbrace{(2\pi -{\beta}/{\eta_y})}_{:= h_1(\beta) } UU^\top 
\end{split}
\end{align*}

For the second term, using $B = T^\top U^\top $ and $G = U\Sigma U^\top$, we obtain
\begin{align*}
M_3   
& = \frac{\beta^2}{c_1\eta_\lambda^2} U \left[ \frac{\beta^4}{\alpha^2}TT^\top + \eta_y^2 \Sigma TT^\top \Sigma
\right.\\
&\phantom{=\;\;}\qquad\qquad \qquad\quad \left.
- \frac{\eta_y\beta^2}{\alpha}(\Sigma TT^\top +TT^\top \Sigma)\right] U^\top\\
&\preceq    \beta\cdot h_2(\alpha,\beta)\cdot UU^\top 
\end{align*}
where 
\begin{align*}
h_2(\alpha,\beta)&:= \frac{\beta}{c_1\eta_\lambda^2}\left(   
\frac{\beta^4}{\alpha^2}||TT^\top|| 
+ \eta_y^2 ||\Sigma TT^\top \Sigma|| \right.\\
&\phantom{=\;\;}\left. \qquad\qquad+ \frac{\eta_y\beta^2}{\alpha}||\Sigma TT^\top +TT^\top \Sigma||
\right)
\end{align*}

For the third term, using $B = T^\top U^\top $ and $G = U\Sigma U^\top$, we have
\begin{align*}
M_2 &=  (\beta-1)A^\top B -\frac{\beta}{c_1\eta_\lambda^2} (  \eta_xF - \frac{\beta^2}{\alpha} I)A^\top B(\frac{\beta^2}{\alpha}I-\eta_y G) \\
& = [ (\beta-1)A^\top  T^\top \\
&\quad\underbrace{ -\frac{\beta}{c_1\eta_\lambda^2} (  \eta_xF - \frac{\beta^2}{\alpha} I)A^\top (\frac{\beta^2}{\alpha}T^\top -\eta_y BU\Sigma) ]}_{:=H_1(\alpha,\beta)}\cdot U^\top
\end{align*}
and thus
\begin{align*}
M_2^\top M_2 
& = U\cdot H_1^\top H_1 \cdot U^\top \preceq h_3(\alpha,\beta)\cdot UU^\top
\end{align*}
where 
\begin{align*}
&h_3(\alpha,\beta): = \\ &\left[ ||TA||  + \frac{\beta}{c_1\eta_\lambda^2} ||A||(  \eta_x\ell + \frac{\beta^2}{\alpha}) (\frac{\beta^2}{\alpha}||T|| +\eta_y|| BU\Sigma||)   \right]^2 
\end{align*}
Here, we use $|\beta -1|\leq 1$ since $0<\beta<1$.

In summary, we obtain
\begin{align*}
{Q}_1^*(\bz) &= Q_y-M_3 - \frac{1}{c_2} M_4^\top M_4\\
& \succeq \underbrace{\left[  \beta\left( h_1(\beta)-h_2(\alpha,\beta)\right) -\frac{h_3(\alpha,\beta)}{c_2}   \right]}_{:=h_4(\alpha,\beta)}\cdot U U^\top
\end{align*}

For sufficiently large $\alpha$ and sufficiently small $\beta$,  $h_3$ is bounded.
 Since $c_2$ is dominated by $\alpha$, we can set $\alpha$ large enough such that $ \frac{h_3}{c_2}\leq \beta^2$ for any fixed $\beta$, 
 then select $\beta$ positively small enough such that $h_1 -h_2-\beta >0 $.
Thus we have $h_4(\alpha,\beta)>0$ and ${Q}_1^*(\bz)\succeq 0$ for any $\bz\in \R^{n+m+k}$.

\subsection{Proof of theorem \ref{local}}\label{pf-local}

Let $H_g^* = \nabla^2 g(\by^*) $ be the Hessian matrix of $g(\by)$ at $\by^*$
and ${D}(\delta) = \{\bz \, | \, || \bz -\bz^*||\leq \delta \}$ be the $\delta$-neighbor of $\bz^*$.
Using Taylor's expansion, we have 
\begin{align*}
\bm{\theta}(\by) := \nabla g(\by) - \left(\nabla g(\by^*) + H_g^*\cdot \Delta \by  \right) \sim O(||\Delta \by||^2)
\end{align*}
which means that there exist positive constants $m_0$ and 
$\delta$ such that  we have 
$||\bm{\theta}(\by)|| \leq m_0 \cdot||\Delta \by||^2 $ for any $\bz\in {D}(\delta) $. Since
\begin{align*} 
\begin{split}
\frac{d \bz}{d t} 
&=\underbrace{\begin{bmatrix}
-  \eta_x  F(\bx) & 0 & - \eta_x A^\top \\ 0 & - \eta_y H_g^* & - \eta_y B^\top \\  \eta_\lambda A &\eta_\lambda B & 0
\end{bmatrix}}_{:= W_0(\bx)} \Delta \bz -\eta_y  \underbrace{\begin{bmatrix}
0\\ \bm{\theta}(\by) \\0
\end{bmatrix}}_{:=\hat{ \bm{\theta}}(\by)}
\end{split}
\end{align*}
we have 
\begin{align*}
\frac{d V_2(\bz)}{dt}&  = \left[  W_0(\bx) \Delta\bz - \eta_y \hat{ \bm{\theta}}(\by)\right]^\top P_2 \Delta \bz\\
&\qquad\qquad\quad +\Delta \bz^\top P_2 \left[ W_0(\bx) \Delta\bz - \eta_y \hat{ \bm{\theta}}(\by) \right]  \\
& = \Delta \bz^\top  W_0(\bx)^\top P_2 \Delta \bz +\Delta \bz^\top P_2 W_0(\bx)\Delta \bz \\
&\quad-2\alpha \Delta \by^\top  \bm{\theta}(\by)  + 2 \frac{\eta_y}{\eta_\lambda} \beta \Delta \bl^\top B  \bm{\theta}(\by)
\end{align*} 
where $V_2(\bz)$ and $P_2$ are defined as (\ref{V2}) and (\ref{P2}) respectively.

By the proof of theorem \ref{thm-psd} and condition (\ref{loo}), there exists $\tau_0 >0$ such that for any $\bz\in D(\delta)$
\begin{align*}
\frac{d V_2(\bz)}{dt}& \leq -\tau_0  \Delta \bz^\top  P_2 \Delta \bz -2\left[ \alpha \Delta \by^\top  -  \frac{\beta\eta_y }{\eta_\lambda} \Delta \bl^\top B   \right] \bm{\theta}(\by)  \\
&\leq -\tau_0  \Delta \bz^\top  P_2 \Delta \bz + 2m_0\left( \alpha+\frac{\beta\eta_y }{\eta_\lambda}||B|| \right) ||\Delta \bz||^3
\\
& \leq  -\frac{\tau_0}{2}  \Delta \bz^\top  P_2 \Delta \bz - \left( \frac{\tau_0}{2}\sigma_{\min} - m_1 ||\Delta \bz||     \right)||\Delta \bz||^2
\end{align*}
where $m_1: =2m_0\left( \alpha+\frac{\beta\eta_y }{\eta_\lambda}||B|| \right) $ and $\sigma_{\min}$ is the minimal eigenvalue of $P_2$.  

Taking $\delta $ sufficiently small such that $\delta \leq \frac{\tau_0}{2m_1}\sigma_{\min}$, we have 
\begin{align*}
\frac{d V_2(\bz)}{dt}& \leq -\frac{\tau_0}{2}  \Delta \bz^\top  P_2 \Delta \bz = -\frac{\tau_0}{2}  V_2(\bz)
\end{align*}

Since $\bz(t)$ asymptotically converges to $\bz^*$, there exists a time $t_\delta\geq 0$ such that $||\Delta \bz||\leq \delta$ for any $t\geq t_\delta$. Hence, theorem \ref{local} is proved.

\subsection{Proof of Lemma \ref{Psd}}\label{pf-Psd}
We use the Schur complement argument twice to prove lemma \ref{Psd}. Firstly, consider the Schur complement of the block $\frac{\alpha}{\eta_x}I$ in matrix $P_1$ (\ref{P1}), which is denoted as $\hat{P}_1$ 
\begin{align*}
\hat{P}_1 &= \begin{bmatrix}
\frac{\alpha}{\eta_y}UU^\top &  - \frac{\beta}{\eta_\lambda}B^\top \\ -\frac{\beta}{\eta_\lambda}B & \frac{\alpha}{\eta_\lambda} I - \frac{\eta_x}{\alpha\eta_\lambda^2} AA^\top
\end{bmatrix} \\
&\succeq\begin{bmatrix}
\frac{\alpha}{\eta_y}UU^\top &  - \frac{\beta}{\eta_\lambda}B^\top \\ -\frac{\beta}{\eta_\lambda}B &  s_1(\alpha)\cdot I 
\end{bmatrix} := \bar{P}_1
\end{align*}
where $s_1(\alpha):=\frac{\alpha}{\eta_\lambda}-\frac{\eta_x}{\alpha\eta_\lambda^2} \kappa_2$. 
Then consider the Schur complement of the block $s_1(\alpha)\cdot I$ in matrix $\bar{P}_1$, which is 
\begin{align*}
\tilde{P}_1 &=\frac{\alpha}{\eta_y}UU^\top - \frac{1}{s_1(\alpha)}\cdot \frac{\beta^2}{\eta_\lambda^2 } B^\top B
\\
& = U \underbrace{\left[\frac{\alpha}{\eta_y}I -\frac{1}{s_1(\alpha)}\cdot  \frac{\beta^2}{\eta_\lambda^2} TT^\top \right]}_{:=S_1(\alpha,\beta)}  U^\top 
\end{align*}
For $\alpha>0$, $s_1(\alpha)$ is strictly increasing in $\alpha$ and $s_1(\alpha)\to +\infty$ as $\alpha\to +\infty$. Hence, $S_1(\alpha,\beta)\succ 0$ when we set parameter $\alpha$ large enough and $\beta$ positively small enough.
 By the Schur complement theorem, we have $\tilde{P}_1\succeq 0$, $\bar{P}_1\succeq 0$, $\hat{P}_2\succeq 0$ and eventually ${P}_1\succeq 0$.

Denote $\Delta\bx = \bx - \bx^*$, $\Delta\by = \by - \by^*$ and $\Delta\bl = \bl - \bl^*$. Then for the  Lyapunov function $V_1(\bz)$ (\ref{V1}),  we have 
\begin{align*}
V_1(\bz)
&= \frac{\alpha}{\eta_x} ||\Delta\bx||^2 + \frac{\alpha}{\eta_y}||U^\top \Delta\by||^2 + \frac{\alpha}{\eta_\lambda} ||\Delta\bl||^2\\&\qquad 
 + \frac{2}{\eta_\lambda}\Delta \bx^\top A^\top \Delta \bl -\frac{2\beta}{\eta_\lambda} \Delta \by^\top B^\top \Delta \bl \\
= & ||\frac{1}{\eta_\lambda}A \Delta \bx + \D \bl||^2+ ||\frac{\beta}{\eta_\lambda} U^\top \D \by -T\D \bl||^2\\
& \qquad
+\D\bx^\top \left( \frac{\alpha}{\eta_x} I -\frac{1}{\eta_\lambda^2}A^\top A  \right)\D\bx \\
&\qquad+ \left( \frac{\alpha}{\eta_y} -\frac{\beta^2}{\eta_\lambda^2}\right) ||U^\top \Delta\by||^2 \\
&\qquad + 
\D \bl^\top \left( (\frac{\alpha}{\eta_y}-1)I-T^\top T  \right)\D \bl
\end{align*}
Set parameter $\alpha$ sufficiently large and $\beta$  positively small enough such that 
\begin{align*}
{\alpha}/{\eta_x}  >{\kappa_2}/{\eta_\lambda^2},\ \ 
 {\alpha}/{\eta_y} > {\beta^2}/{\eta_\lambda^2}, \ \ {\alpha}/{\eta_\lambda}-1> ||T^\top T||,
\end{align*}
Then 
$
V_1(\bz)=0 \Longleftrightarrow
\D \bx =\bm{0},\  \D \bl =\bm{0}, U^\top \D \by =\bm{0}
$
and 
\begin{align*}
U^\top \D \by =\bm{0} \Longleftrightarrow B\D \by =\bm{0}, G\D \by =\bm{0}
\end{align*}
due to $ker(G)\cap ker(B) = span(\bu_{l+r+1},\cdots,\bu_m)$.

\subsection{Proof of Lemma \ref{Ppd}} \label{pf-Ppd}
Consider the Schur complement of the block $\frac{\alpha}{\eta_x}I$ in matrix $P_2$ (\ref{P2}), which is 
\begin{align*}
\hat{P}_2 = \begin{bmatrix}
\frac{\alpha}{\eta_y}I &  - \frac{\beta}{\eta_\lambda}B^\top \\ -\frac{\beta}{\eta_\lambda}B & \frac{\alpha}{\eta_\lambda} I - \frac{\eta_x}{\alpha\eta_\lambda^2} AA^\top
\end{bmatrix} 
\end{align*}
Then consider the Schur complement of the block $\frac{\alpha}{\eta_y}I$ in matrix $\hat{P}_2$, which is 
\begin{align*}
\bar{P}_2 &= \frac{\alpha}{\eta_\lambda} I - \frac{\eta_x}{\alpha\eta_\lambda^2} AA^\top - \frac{\eta_y\beta^2}{\alpha\eta_\lambda^2} BB^\top \\
&\succeq \underbrace{\left( {\frac{\alpha}{\eta_\lambda} - \frac{\eta_x}{\alpha\eta_\lambda^2} \kappa_2- \frac{\eta_y\beta^2}{\alpha\eta_\lambda^2} ||BB^\top|| }\right)}_{ p_1(\alpha,\beta)} \cdot I 
\end{align*}
It is easy to check that $p_1(\alpha,\beta)>0$ when we set parameter $\alpha$ large enough and $\beta$ positively small enough, thus $\bar{P}_2\succ 0$. By the Schur complement theorem, we have  $\hat{P}_2\succ 0$ and eventually ${P}_2\succ 0$.

Since $P_2\succ 0$, we have 
$
V_2(\bz) =0 \Longleftrightarrow \bz =\bz^*
$.

\subsection{Proof of Proposition \ref{eps-q} and Proposition \ref{eqs-ge}} \label{pf-epsq}

Let $\hat{\bz} := \left[ \hat{\bx}^{\top}, \hat{\by}^{\top}, \hat{\bl}^{\top}\right]^\top$ be another equilibrium point of PDGD (\ref{pdg}).
	By the definition, we have 
	\begin{subequations}
		\begin{align}
		\nabla f(\hat{\bx}) - \nabla f(\bx^*) &= -A^\top (\hat{\bl}-\bl^*) \label{ep-x}\\
		\nabla g(\hat{\by}) - \nabla g(\by^*) &= -B^\top (\hat{\bl}-\bl^*) \label{ep-y}\\
		\bm{0}&=	A(\hat{\bx}-\bx^*) + B(\hat{\by}-\by^*)  \label{ep-l}
		\end{align}
	\end{subequations}
	Multiply $(\hat{\bx}-\bx^*)^\top$, $(\hat{\by}-\by^*)^\top$ and $(\hat{\bl}-\bl^*)^\top$ to the both sides of (\ref{ep-x}),  (\ref{ep-y}) and (\ref{ep-l}) respectively and sum them up, we obtain 
	\begin{align*}
	(\nabla &f(\hat{\bx}) - \nabla f(\bx^*))^\top (\hat{\bx}-\bx^*)\\ &\qquad+ (\nabla g(\hat{\by}) - \nabla g(\by^*))^\top (\hat{\by}-\by^*) =\bm{0}
	\end{align*}
	Due to the strong convexity of $f(\bx)$ and convexity of $g(\by)$, we have 
		\begin{align}  \label{x-x}
		\hat{\bx}=\bx^*, \ 
		 ( g(\hat{\by})-\nabla g(\by^*))^\top (\hat{\by}-\by^*)=\bm{0}  
		\end{align}

	Since $A$ is of full row rank (assumption \ref{aa}), we  have $\hat{\bl}=\bl^*$ due to (\ref{ep-x}) and (\ref{x-x}). By  (\ref{ep-l}), we obtain $B\hat{\by}=B\by^*$. By (\ref{ep-y}), we have $\nabla g(\hat{\by})=\nabla g(\by^*)$, which is $G\hat{\by} = G\by^*$ in (\ref{eqs-qua}). Hence, proposition \ref{eps-q} is proved.

	For proposition \ref{eqs-ge}, by assumption \ref{ge-psd} we further have $$||\hat{\by}-\by^*||^2\leq 0 \Longrightarrow \hat{\by}=\by^* $$ 
Hence, the equilibrium point $(\bx^*,\bl^*,\by^*)$ is unique.

\subsection{Proof of Proposition \ref{syn}} \label{pf_syn}

	From (\ref{dk-x}), we have 
	\begin{align*}
	&\bx_{i+1} -\bx^* =  \bx_i-\bx^*\\ 
	&\qquad   -\nu_x (\nabla f(\bx_i) -\nabla f(\bx^*) + A^\top (\bl_i-\bl^*))
	\\
	\Longrightarrow\ & ||A^\top (\bl_i-\bl^*)|| \leq  ||\nabla f(\bx_i) -\nabla f(\bx^*) ||\\
	& \qquad\qquad +\frac{1}{\nu_x}\left(||\bx_i-\bx^*|| + ||\bx_{i+1} -\bx^*||\right)  \\
	&\qquad\qquad \qquad\ \ \leq  \left(\ell+ \frac{2}{\nu_x} \right)||\bx_i-\bx^*|| \\
	\Longrightarrow\ & ||\bl_i-\bl^*  || \leq \frac{\ell+ {2}/{\nu_x}}{\sqrt{\kappa_1}} \cdot||\bx_i-\bx^*|| \leq c_\lambda \cdot \vartheta^i
	\end{align*}
	where $c_\lambda :=\frac{c_x}{\sqrt{\kappa_1}} (\ell+ \frac{2}{\nu_x} )$.

	Similarly, from (\ref{dk-l}), we have
	\begin{align*}
   & ||B(\by_i-\by^*)|| \leq  ||A(\bx_i-\bx^*)||\\
	&\qquad\qquad +\frac{1}{\nu_\lambda}(||\bl_{i+1} -\bl^*||+|| \bl_{i} -\bl^*||)  \\
	&\quad \leq ||A||\cdot||\bx_i-\bx^*||+\frac{2}{\nu_\lambda}||\bl_i-\bl^*||\leq c_y \cdot \vartheta^i
	\end{align*}
	where $c_y := c_x\cdot||A||+\frac{2c_\lambda}{\nu_\lambda}$.
	
	From (\ref{dk-y}), we have 
	\begin{align*}
	& ||B\left(\nabla g(\by_i)-\nabla g(\by^*)\right)||\leq ||BB^\top||\cdot||\bl_i-\bl^*|| \\
&\qquad \qquad\qquad\qquad	+\frac{2}{\nu_y}||B(\by_i-\by^*)|| \leq c_g \cdot \vartheta^i
	\end{align*}
	where $c_g :=  c_\lambda\cdot||BB^\top||+\frac{2c_y}{\nu_y}$.

\end{document}